\documentclass[twocolumn,showpacs,superscriptaddress,preprintnumbers,amsmath,amssymb,longbibliography]{revtex4-1}
\usepackage{graphicx}
\usepackage{dcolumn}
\usepackage{bm}
\usepackage{amsmath}
\usepackage{color}

\begin{document}

\title{Onset and cessation of motion in hydrodynamically sheared granular beds}

\author{Abram H. Clark}
\affiliation{Department of Mechanical Engineering and Materials Science, Yale University, New Haven, Connecticut 06520, USA}
\author{Mark D. Shattuck}
\affiliation{Benjamin Levich Institute and Physics Department, The City College of the City University of New York, New York, New York 10031, USA}
\author{Nicholas T. Ouellette}
\affiliation{Department of Civil and Environmental Engineering, Stanford University, Stanford, California 94305, USA}
\affiliation{Department of Mechanical Engineering and Materials Science, Yale University, New Haven, Connecticut 06520, USA}
\author{Corey S. O'Hern}
\affiliation{Department of Mechanical Engineering and Materials Science, Yale University, New Haven, Connecticut 06520, USA}
\affiliation{Department of Physics, Yale University, New Haven, Connecticut 06520, USA}
\affiliation{Department of Applied Physics, Yale University, New Haven, Connecticut 06520, USA}

\begin{abstract}
We performed molecular dynamics simulations of granular beds driven by
a model hydrodynamic shear flow to elucidate general grain-scale
mechanisms that determine the onset and cessation of sediment
transport. By varying the Shields number (the nondimensional shear
stress at the top of the bed) and particle Reynolds number (the
ratio of particle inertia to viscous damping), we explore how
variations of the fluid flow rate, particle inertia, and fluid
viscosity affect the onset and cessation of bed motion. For low to
moderate particle Reynolds numbers, a critical boundary separates
mobile and static states. Transition times between these states
diverge as this boundary is approached both from above and below. At
high particle Reynolds number, inertial effects become dominant, and
particle motion can be sustained well below flow rates at which
mobilization of a static bed occurs. We also find that the onset of
bed motion (for both low and high particle Reynolds numbers) is
described by Weibullian weakest-link statistics, and thus is crucially
dependent on the packing structure of the granular bed, even deep
beneath the surface. 
\end{abstract}

\date{\today}


\maketitle

\section{\label{sec:intro}Introduction}

Fluid flowing laterally over a granular bed exerts shear stress on the
grains. This occurs in many natural settings and industrial
applications, such as sediment transport in
riverbeds~\cite{einstein1950,charru2013} and slurries
pipes~\cite{doron87,capecelatro2013}. The ratio of the shear stress
exerted by the fluid on the top of the bed to the buoyancy-corrected
particle weight is known as the Shields number
$\Theta$~\cite{shields1936}. For small $\Theta$, no grain motion
occurs; at sufficiently large $\Theta$, however, grains can be
entrained by the
flow~\cite{wilberg1987,buffington1997,charru2004,ouriemi2007,beheshti2008,dey2008,lajeunesse2010,dersken2011,hong2015,houssais2015}. Despite
decades of research, the nature of the transition between static and
mobilized granular beds is not well understood. The geometric
structure of the contact network in the bed determines its mechanical
strength~\cite{silbert2002,peyneau08,schreck2010}. Bed mobilization is
also strongly affected by the complex and unsteady fluid flow above
the bed, as well as how strongly the fluid flow couples to the grains, as quantified by the particle
Reynolds number $Re_p$~\cite{shields1936,buffington1997,dey2008,beheshti2008} that measures
how quickly grains equilibrate to the fluid flow. Weak
stresses applied to the interior of the bed by fluid flowing through
the pore spaces between grains may also play a role in bed
mobilization~\cite{rose1945,beavers1967,scheidegger1974}. Although empirical
hydraulic models capture some important aspects of sediment transport
problems~\cite{charru2004}, there is at present no fundamental understanding of the
relative contributions of these effects on the onset and cessation of grain motion.

In this paper, we study a simplified model of a fluid-driven granular
bed to clarify the essential physics at the onset of bed motion. In
particular, we seek to understand the nature of the mobile-to-static
and static-to-mobile transitions as a function of $\Theta$ and $Re_p$
and to predict the parameter regime where hysteresis, defined as a
finite difference between $\Theta_0$, above which a static bed will
begin to move, and $\Theta_c$, below which a mobile system will come
to rest, occurs.

We performed molecular dynamics (MD) simulations of a two-dimensional
(2D) system composed of frictionless disks subjected to a simplified
fluid flow that decays from a large value above the bed to a small value inside the bed. Although our model is highly
simplified, with, for example, no explicit unsteadiness in the flow or
friction between the grains, we find that $\Theta_c(Re_p)$ from the
simulations is consistent with the behavior obtained from a large
collection of experiments on sediment transport~\cite{buffington1997,dey2008,beheshti2008}.
In particular, we find plateau values $\Theta_c^{l}$ and $\Theta_c^h$
at low and high $Re_p$, with $\Theta_c^l > \Theta_c^h$, and an
intermediate $Re_p$ regime that connects the two limiting values. In
the low $Re_p$ limit, there is a sharp transition at $\Theta_c$
between mobile and static beds in the infinite-time and
infinite-system-size limits with no hysteresis. In the large $Re_p$
limit, we find significant hysteresis, since particle inertia can
sustain motion well below the $\Theta_0$ at which bed motion is
initiated. We also find that the onset of bed motion at
$\Theta>\Theta_c$ for low $Re_p$ and $\Theta>\Theta_0$ for high $Re_p$
depends strongly on system size and exhibits weakest-link
statistics~\cite{weibull1939,weibull1951}. Thus, the onset of bed
motion in our system depends on the bed packing structure, even deep
beneath the surface.

\section{\label{sec:model}Details of the model}

\begin{figure}
\raggedright (a) \\ \centering \includegraphics[trim=10mm 5mm 10mm 5mm, clip, width=.7\columnwidth]{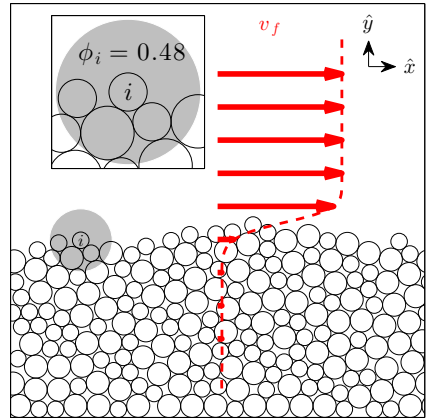}
\\ \raggedright (b) \\ \centering \includegraphics[trim=10mm 5mm 10mm 5mm, clip, width=.7\columnwidth]{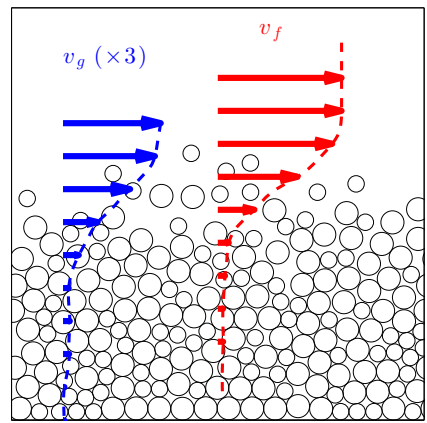}
\caption{(Color online) Layer-averaged fluid velocity $v_f$
versus depth for static (a) and mobile (b) beds. Grains are subjected to a fluid drag force and a gravitational
force $-mg'\hat{y}$. The gray circle (left panel) defines the area
used to calculate the local packing fraction $\phi_i$ near the $i$th
particle, which determines the local fluid velocity. We also show
the layer-averaged grain velocity $v_g$ for a mobile bed in the
right panel.}
\label{fig:cartoon}
\end{figure}

We study a domain of width $W$ that contains $N/2$ large and $N/2$
small disks with diameter ratio $1.4$. There is no upper boundary, and
the lower boundary is rigid with infinite friction so that the
horizontal velocities of all particles touching it are set to zero. We
use periodic boundary conditions in the horizontal direction. The
total force on each particle is given by the vector sum of contact
forces from other particles, a gravitational force, and a
Stokes-drag-like force from a fluid that moves horizontally:
\begin{equation}
m_i\vec{a}_i=\sum_j \vec{F}^c_{ij} - m_ig'\hat{y} + B_i[v_0f (\vec{r})\hat{x}-\vec{v}_i ].
\label{eqn:force-law}
\end{equation} 
Here, $m_i\propto D_i^2$ is the particle mass, $D_i$ is the diameter
of particle $i$, $\vec{v}_i$ and $\vec{a}_i$ are the velocity and
acceleration of particle $i$, $m_ig'$ is the buoyancy-corrected
particle weight, $B_i\propto D_i$ sets the drag on disk $i$, $v_0$ is
a characteristic fluid velocity at the surface of a static bed, and
$f(\vec{r})$ is the fluid velocity at $\vec{r}$. $\vec{F}^c_{ij}=
K\left(1-\frac{r_{ij}}{D_{ij}}\right)\theta\left(1-\frac{r_{ij}}{D_{ij}}\right)\hat{r}_{ij}$
is the pairwise repulsive contact force on disk $i$ from disk $j$,
where $K$ is the particle stiffness, $r_{ij}$ is the separation
between the centers of the particles, $D_{ij}=(D_i + D_j)/2$,
$\hat{r}_{ij}$ is the unit vector connecting their centers, and
$\theta$ is the Heaviside step function. $f(\vec{r})$, the fluid
velocity profile, varies smoothly from a large value above the bed to
a small value inside the bed. We choose a form that depends only on
the local packing fraction $\phi_i$: $f(\phi_i)=e^{-b(\phi_i-0.5)}$
where $b$ controls the ratio of the magnitude of the fluid flow above
and inside the bed. $\phi_i$ is calculated in a small circular region
with diameter $D_i+2D_l$ around each particle, as shown in
Fig.~\ref{fig:cartoon}(a). We note that $f=1$ for $\phi_i=0.5$, a
typical value at the bed surface. See Section \ref{sec:protocol-A} for force profiles
in static and mobile states.

Three nondimensional numbers govern the behavior of
Eq.~\eqref{eqn:force-law}. We set the nondimensional stiffness
$\frac{K}{m g'}>3\times 10^3$ to be sufficiently large that increasing
it has no effect on our results. The other two nondimensional
parameters can be written as
\begin{align}
\Theta &= \frac{Bv_0}{mg'} \\
\Gamma &= \frac{B/m}{\sqrt{g'/D}}.
\label{eqn:nondimen-params}
\end{align}
The Shields number $\Theta$ gives the dimensionless shear force at the
top of a static bed. $\Gamma$ is the ratio of the gravitational
settling time $\tau_s=\sqrt{D/g'}$ to the viscous time scale
$m/B$. Since the particle Reynolds number $Re_p = \frac{v_0 D}{\nu}$,
where $\nu$ is the kinematic viscosity, and the Stokes drag is
proportional to $\rho_f \nu D$, where $\rho_f$ is the fluid density,
the ratio $\frac{\Theta}{\Gamma^2} = \frac{mv_0}{BD} \propto
\frac{\rho_g}{\rho_f}Re_p$ ($\rho_g$ is the mass density of the
grains) compares the inertia of grains entrained in the flow to
the strength of the viscous drag.

To characterize flow onset and cessation in our system, we employed
two protocols. In protocol A, to study the mobile-to-static
transition, we distributed particles randomly throughout the domain
and set a constant value of $\Theta$ for a total time of roughly $10^5
\tau_s$. We consider the bed to be at rest when the maximum net
particle acceleration $a_{\rm max}$ is below a threshold $a_{\rm thresh}$
roughly one order of magnitude smaller than $g'$ and roughly three
orders of magnitude smaller than typical values for a moving bed. In
protocol B, to understand the dynamics of the static-to-mobile
transition, we begin with a static bed from protocol A and slowly
increase $\Theta$ in increments $\Delta\Theta = 0.01\Theta$. If
$a_{\rm max}<a_{\rm thresh}$ after roughly one inter-grain collision
time, then $\Theta$ is increased. If $a_{\rm max}>a_{\rm thresh}$, we
keep $\Theta$ constant until $a_{\rm max}<a_{\rm thresh}$. We
designate a system as mobile under protocol B with slightly different
criteria: $a_{\rm max}/a_{\rm thresh}>10$ and $\bar{v}_g>0.04 V_s$, where $\bar{v}_g$ is the average horizontal velocity of all
grains and
$V_s=\sqrt{g'D}$ is the settling velocity. These thresholds filter out
small rearrangement events, keeping only states with substantial
grain motion.

\section{\label{sec:results}Results and Discussion}

\begin{figure} \centering
\raggedright (a) \\ \includegraphics[trim=0mm 0mm 12mm 0mm, clip, width=\columnwidth]{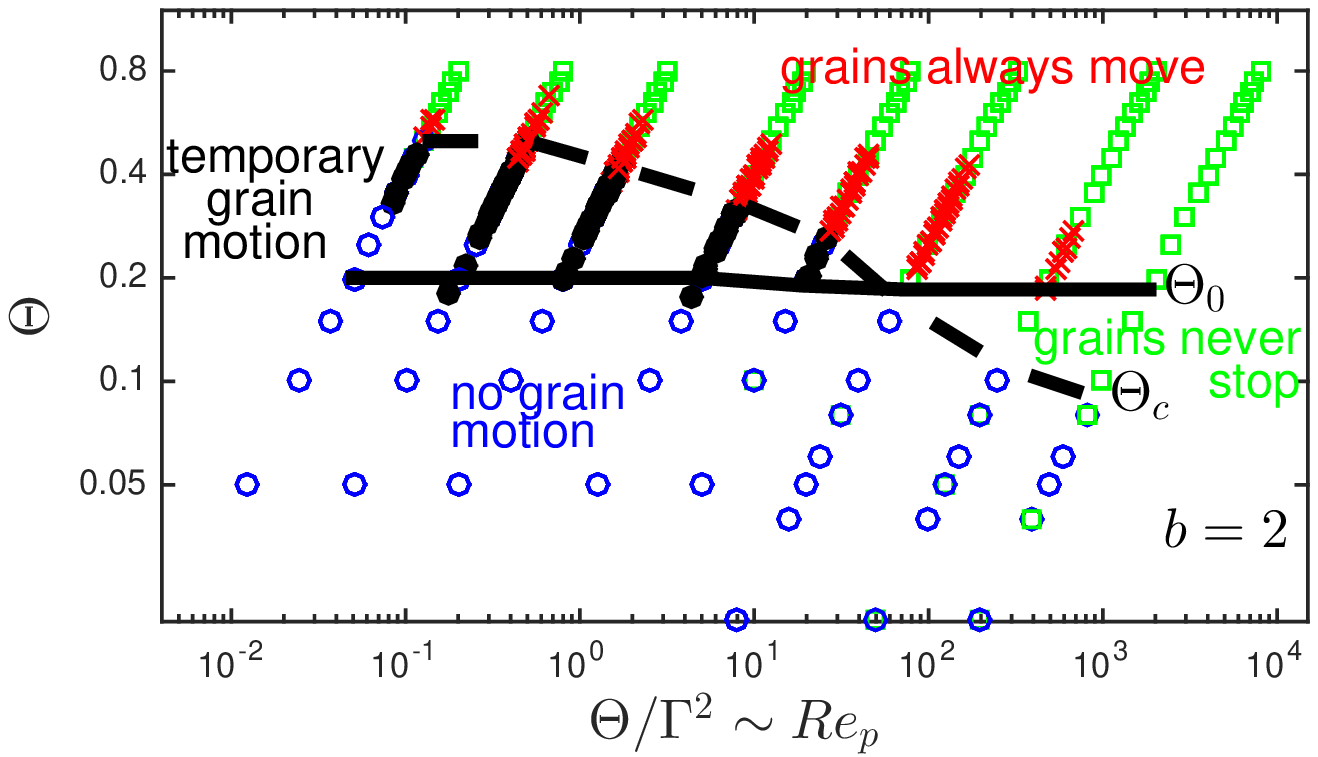}
\\ \raggedright (b) \\ \includegraphics[trim=0mm 0mm 12mm 0mm, clip, width=\columnwidth]{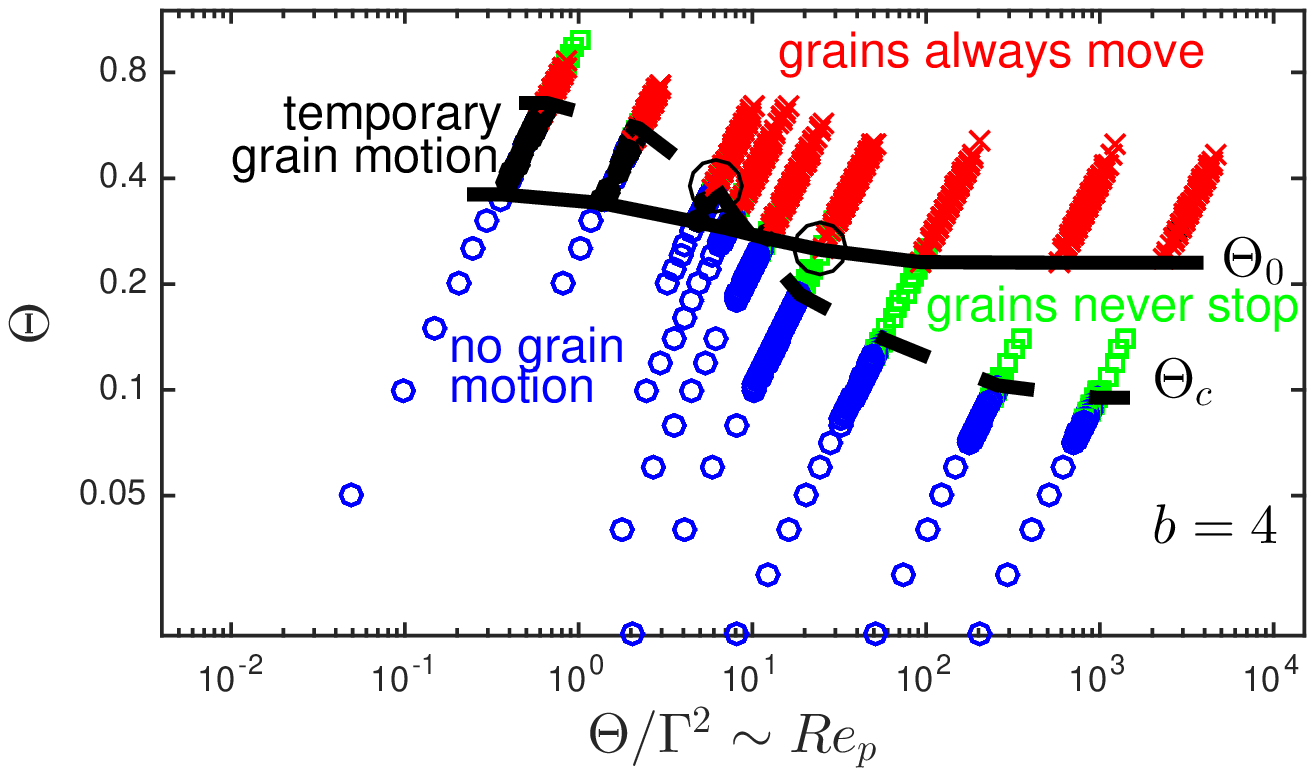}
\\ \raggedright (c) \\ \includegraphics[trim=0mm 0mm 12mm 0mm, clip, width=\columnwidth]{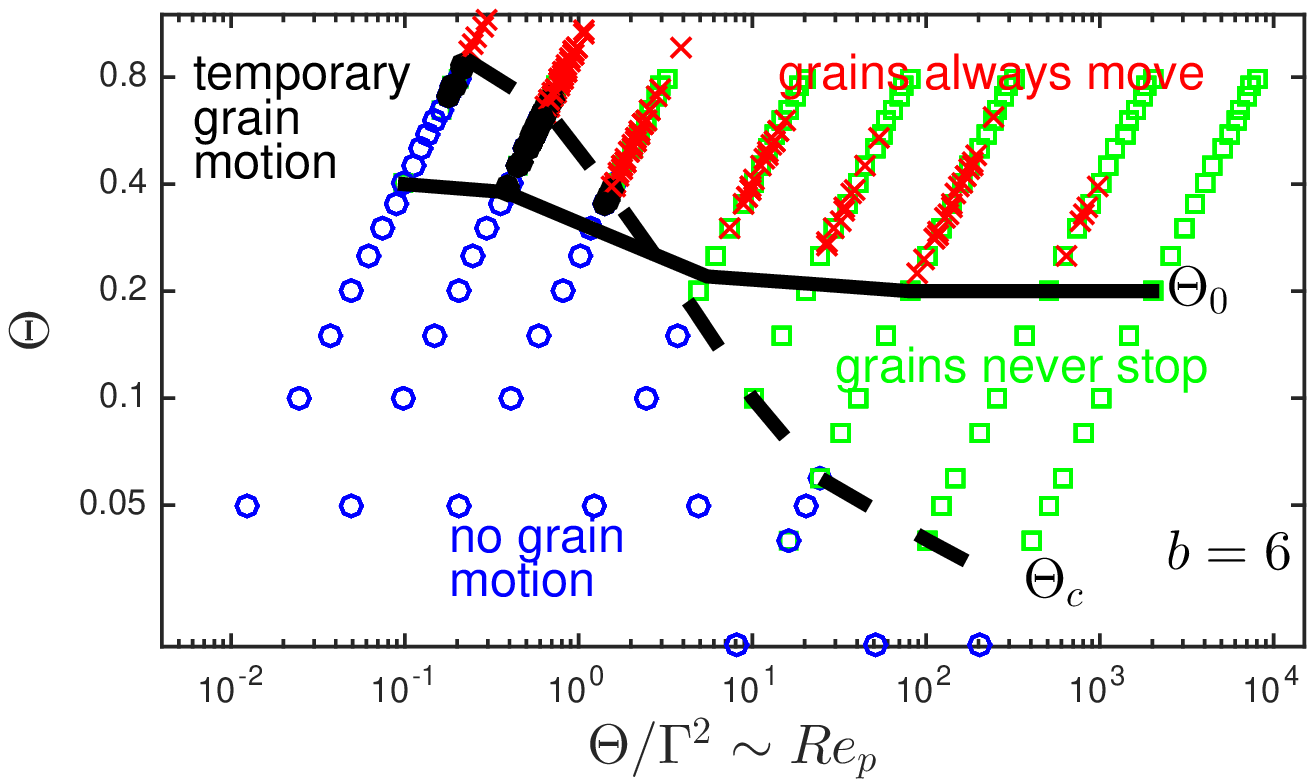}
\caption{(Color online) The flow diagrams, Shields number $\Theta$ versus particle 
Reynolds number $Re_p$, with (a) $b=2$, (b) $b=4$, (c) and $b=6$. Diagonal 
lines of data points correspond to lines of constant $\Gamma$. The symbols show systems that came to rest
(\textcolor{blue}{$\circ$}) or never stopped
(\textcolor{green}{$\square$}) under protocol A, and were
permanently (\textcolor{red}{$\times$}) or temporarily ($\bullet$)
mobile as $\Theta$ was increased under protocol B. The dashed line
shows $\Theta_c$, above which the inertial effects from particles
entrained in the flow lead to sustained motion. The solid line indicates
$\Theta_0$, below which the system will never be mobilized. The two
large black open circles mark the parameter values we study in
Fig.~\ref{fig:Weibull-scaling}.}
\label{fig:phase-diagrams}
\end{figure}

Figure~\ref{fig:phase-diagrams} shows the boundaries between mobile and
static beds as a function of $\Theta$ and $Re_p$ from simulations with
$b=2$, 4, and 6. We find a curve $\Theta_c(Re_p)$ above which the particles are
unable to find a stable packing under protocol A. In the low-$Re_p$
limit near $\Theta_c$, grain motion is highly overdamped, and
particles do not leave the bed. As $Re_p$ is increased, the inertial
effects of mobilized particles striking the bed make finding a stable
configuration more difficult, decreasing $\Theta_c$
significantly. Figure~\ref{fig:phase-diagrams} shows parameter values
where grain motion does (blue circles) and does not (green squares)
stop under protocol A. We then apply protocol B to stopped systems and
find another boundary $\Theta_0(Re_p)$ that specifies when grain flow
can be initiated. For low $Re_p$, $\Theta_0<\Theta_c$, and bed motion
initiated near $\Theta_0$ is temporary. Thus, at low $Re_p$, permanent
grain motion is initiated at $\Theta_c$ in the large system limit, as
we discuss below. However, in the high-$Re_p$ limit where particle inertia is dominant, we observe significant hysteresis: grain motion is initiated at $\Theta_0$, which is well above the value of $\Theta_c$ where mobile particles colliding with the bed can sustain bed motion. We also note that this result is consistent with~\cite{carneiro2011}, where, in simulations of Aeolian transport at high $Re_p$, a significant perturbation or lift force was required near $\Theta_c$ to initiate grain motion. Temporary (filled black circles) and permanent (red crosses) motion under protocol B are also marked in Fig.~\ref{fig:phase-diagrams}. The basic nature of the flow diagram is insensitive to variations in $b$, although the numerical values for $\Theta_0(Re_p)$ and $\Theta_c(Re_p)$ change. We note the similarities of $\Theta_c(Re_p)$ in Fig.~\ref{fig:phase-diagrams} to the experimental and observational data compiled in~\cite{buffington1997,dey2008,beheshti2008}, even though our model is highly simplified. Both display a plateau in the onset value of $\Theta$ at low $Re_p$, a decrease in the onset value at moderate $Re_p$, and a lower plateau value at high $Re_p$.

The $b$ parameter sets the ratios between the fluid velocity above the bed ($v_a$), at the top of a static bed ($v_0$), and in the bulk of the bed ($v_b$) where grains are packed densely (at $\phi_i \approx 0.84$). If a grain is well above the bed, $\phi_i\approx 0.1$ (since the grain itself contributes to the local packing fraction), so the ratio of the fluid velocity for this grain to the fluid velocity at the top of an otherwise static bed (where $\phi_i \approx 0.5$ and thus $f=1$) is $v_a/v_0 \approx e^{0.4b}$. Table~\ref{tbl:b-dependence} shows the ratios $v_a/v_0$, $v_0/v_b$, and $v_a/v_b$ for $b=2$, 4, and 6.
\begin{table}
    \caption{The flow velocity ratios at different heights and the particle Reynolds number when $\Theta_0=\Theta_c$. $v_a$, $v_0$, and $v_b$ are the fluid velocities above the bed, at the surface of the bed, and in the interior of the bed, respectively.}
\centering
    \vspace{0.2in}
    \begin{tabular}{|c|c|c|c|c|}
    \hline
      $b$ & $v_a/v_0$ & $v_0/v_b$ & $v_a/v_b$ & $Re_p$ when $\Theta_0 = \Theta_c$\\ \hline
    2 & 2.23 & 1.97 & 4.39 & $Re_p \approx 50$ \\ \hline
    4 & 4.95 & 3.89 & 19.3 & $Re_p \approx 10$ \\ \hline
    6 & 11.02 & 7.69 & 84.8 & $Re_p \approx 3$ \\ \hline
 \end{tabular}
    \label{tbl:b-dependence}
\end{table}
Note that these ratios increase with $b$, and that varying $b$ changes all ratios simultaneously. Obviously, a fluid flow model with an additional parameter could decouple the $v_a/v_0$ and $v_0/v_b$ ratios, but the model used here was chosen as a very simple way to interpolate between a large fluid velocity above the bed and a small fluid velocity inside the bed. Figure~\ref{fig:phase-diagrams} shows that the flow diagram is qualitatively the same for $b=2$, 4, and 6. $\Theta_0$ is relatively insensitive to $b$, but $\Theta_c$ shifts to smaller $Re_p$ and the gap between the plateau values at large and small $Re_p$ widens with increasing $b$.

\subsection{\label{sec:protocol-A}Protocol A: Mobile-to-static transition}

\begin{figure}
\raggedright (a) \hspace*{33mm} (b) \\ 
\centering \includegraphics[width=0.45\columnwidth]{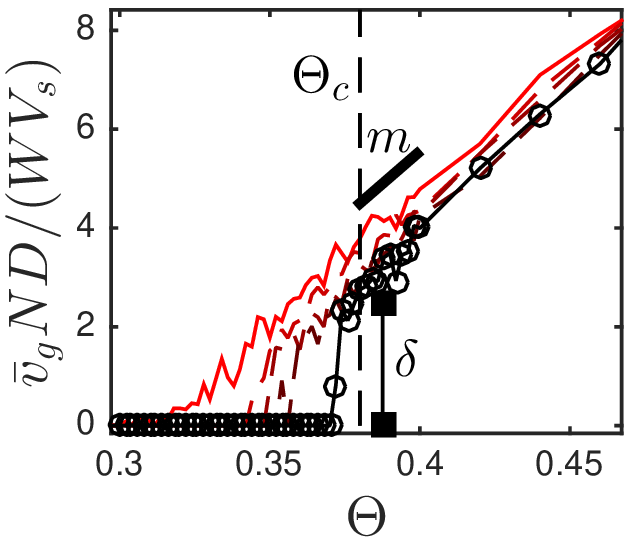}
\includegraphics[width=0.45\columnwidth]{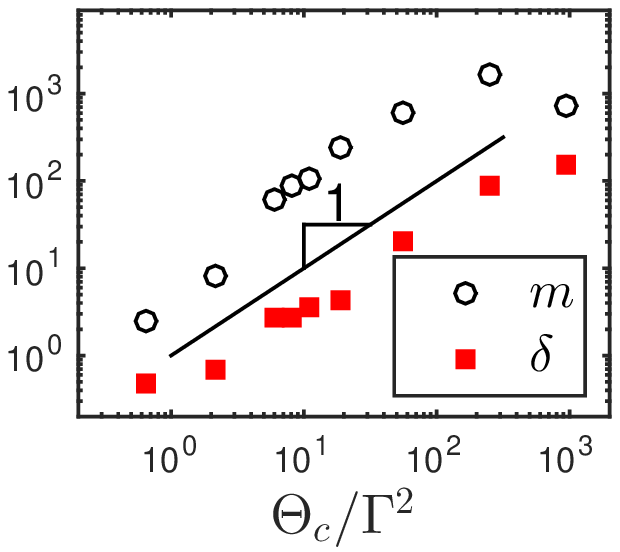}
\\ \raggedright (c) \\ \includegraphics[trim=0mm 0mm 0mm 0mm, clip, width=\columnwidth]{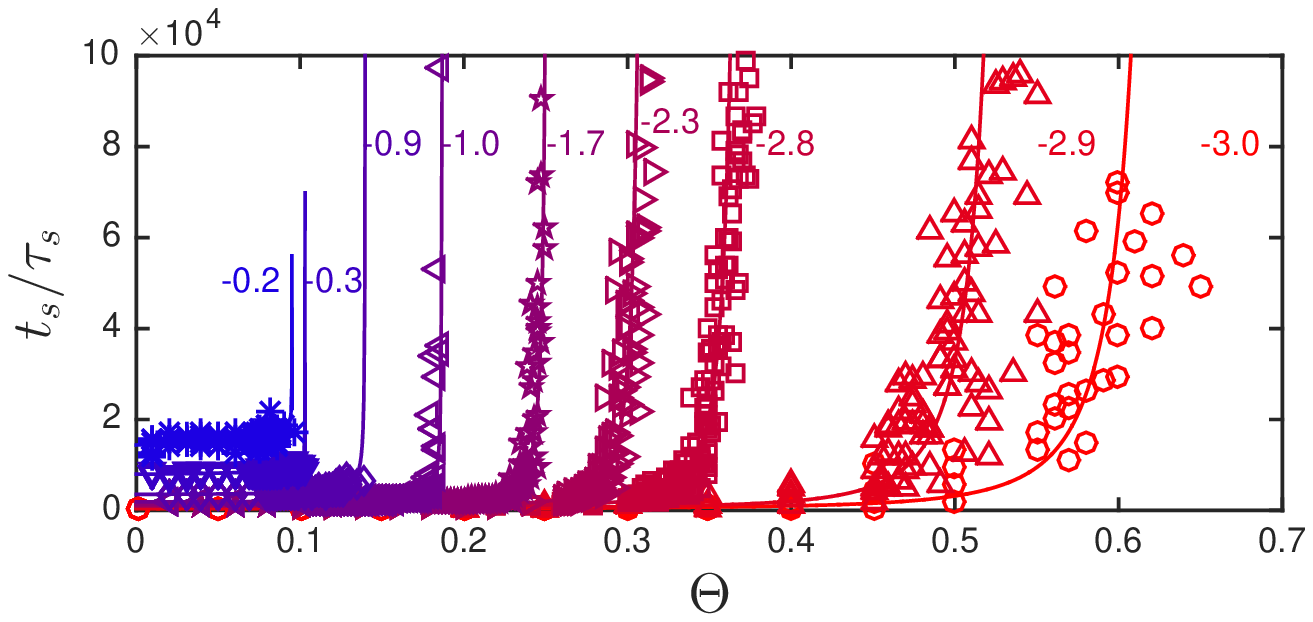}
\\ \raggedright (d) \\  \includegraphics[trim=0mm 0mm 0mm 0mm, clip, width=\columnwidth]{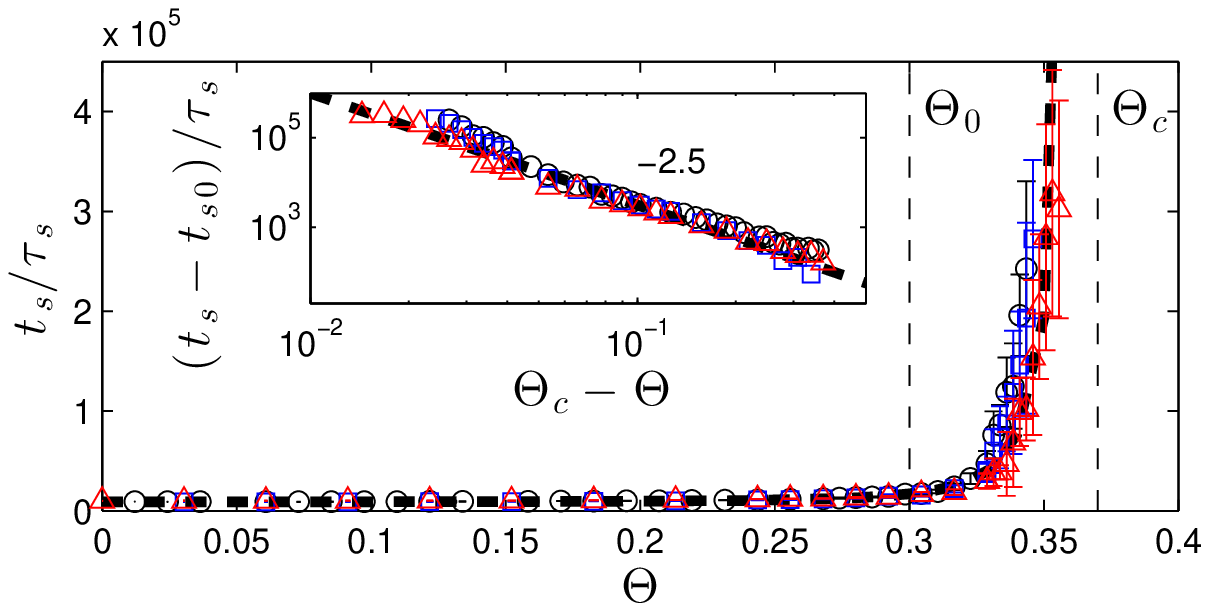}
\caption{(Color online) Data from protocol A near $\Theta_c$. (a) Time evolution of the average grain velocity $\bar{v}_g(t)$ during protocol A. The solid red curve characterizes grain motion at the time for grains to settle under no fluid flow $t_{s0}$, the black curve with open circles marks the end of the simulation, and dashed curves represents intermediate times (where red to black represents increasing time). At long times, we observe a discontinuity in $\bar{v}_g$ with magnitude $\delta$ and slope $m$ at $\Theta_c$. (b) $\delta$ and $m$ scale roughly linearly with $Re_p$. (c) The time $t_s$ required for the grains to come to rest as a function of $\Theta$ for varying $Re_p$ (left to right, blue to red, represents large to small $Re_p$); symbols show $\Gamma=0.01$ ($\ast$), $0.02$ ($\triangledown$), $0.05$ ($\diamond$), $0.1$ ($\triangleleft$), $0.15$ ($\star$), $0.2$ ($\triangleright$), $0.25$ ($\square$), $0.5$ ($\triangle$), and $1$ ($\circ$). The lines show fits of $(t_s-t_{s0})\propto(\Theta_c - \Theta)^{\alpha}$, where the values of $\alpha$ are marked next to each plot. (d) $t_s$ is plotted as a function $\Theta$ for $b=4$, $\Gamma=0.24$, and three system sizes. The symbols correspond to ($W/D$,$N$): black circles (100,800), blue squares (50,400), and red triangles (50,800). Data points show the mean of ten simulations, and error bars give the standard deviation. The inset shows a logarithmic plot of the mean of $t_s-t_{s0}$. The thick dashed line corresponds to $(t_s-t_{s0})\propto(\Theta_c - \Theta)^{-2.5}$.}
\label{fig:protocol-A}
\end{figure}

Figure~\ref{fig:protocol-A} shows data from protocol A near
$\Theta_c$, which we find to be nearly independent of system size, as we discuss below. In Fig.~\ref{fig:protocol-A}(a), we show the
time evolution of $\bar{v}_g$ multiplied by the fill height $ND/W$ and normalized by
$V_s$. At short times (red curve), $\bar{v}_g$ increases linearly with
$\Theta$ and connects continuously to $\bar{v}_g = 0$. This is the
expected relation for frictionless granular systems: no motion occurs
below the yield stress, while above the yield stress the strain rate
increases as a power law in the difference between the applied and
yield stresses~\cite{xu2006}. However, at long times we observe a
discontinuity in $\bar{v}_g$ at $\Theta_c$, as has also been observed
in experiments~\cite{lajeunesse2010} as well as simulations of Aeolian
transport~\cite{carneiro2011} and sheared frictional granular
media~\cite{otsuki2011}. The discontinuity in $\bar{v}_g$ moves toward
$\Theta_c$ as time increases. The size of the discontinuity and the
slope $\bar{v}_g/\Theta$ both scale roughly linearly with $Re_p$, as
shown in Fig.~\ref{fig:protocol-A}(b). For Shields numbers below
$\Theta_c$, the grains settle into a stable packing in a time $t_s$
that diverges as $\Theta\rightarrow\Theta_c$, as shown in
Fig.~\ref{fig:protocol-A}(c).

Figure~\ref{fig:cartoon} shows that mobile grains are confined to a relatively small layer at the top of the bed. So, for fill heights studied here ($ND/W>8$), we find the total flow rate $\bar{v}_g N D/W$ as a function of $\Theta$ and $\Gamma$ is insensitive to the system size (recall, $\bar{v}_g$ is the average velocity of all grains). One might think that the discontinuity in the the flow rate near $\Theta_c$ may disappear in the large-system limit: small systems of grains are able to find a stable configuration, but very large systems will always have a weak spot that does not allow the system to stop. However, this is not the case. Figure~\ref{fig:protocol-A}(d) shows that the behavior near $\Theta=\Theta_c$ is insensitive to the system size. The stopping time as a function of $\Theta$ is shown for three different system sizes, and the three curves are virtually identical. The mean and fluctuations of the stopping time both diverge as a power law as $\Theta\rightarrow\Theta_c$. The data shown is for $\Gamma=0.24$ and $b=4$, where $\Theta_c>\Theta_0$, and the power law exponent is roughly 2.5. As shown in Fig.~\ref{fig:protocol-A}(c), this exponent is roughly 3 in the low $Re_p$ limit, and it decreases with increasing $Re_p$ to less than 1 in the high $Re_p$ limit. $\Theta_0$ is also shown, and it corresponds reasonably well to where the divergence begins, which suggests that $\Theta_c-\Theta_0$ is related to the power law exponent of the divergence.

\begin{figure}
\centering
\raggedright (a) \hspace*{35mm} (b) \\ \centering
\includegraphics[trim=0mm 0mm 5mm 0mm, clip, width=0.49\columnwidth]{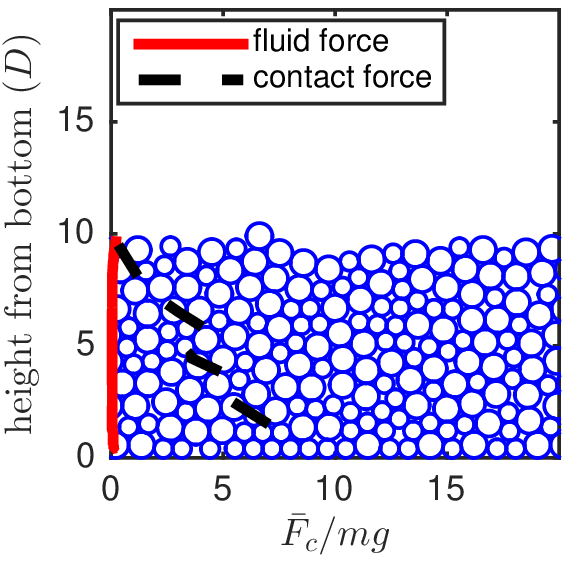}
\includegraphics[trim=0mm 0mm 5mm 0mm, clip, width=0.49\columnwidth]{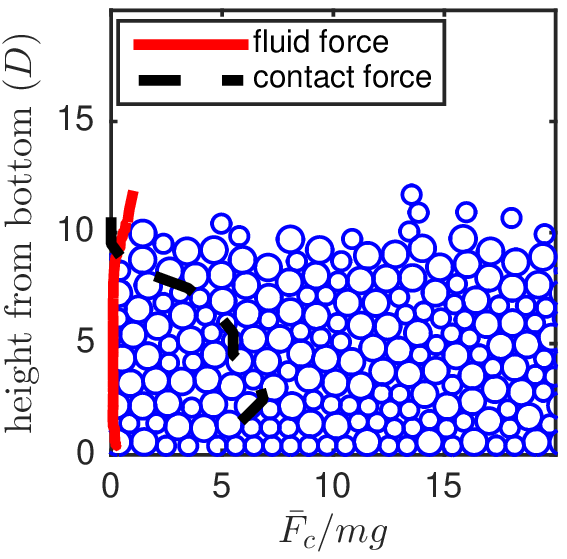}
\\ \raggedright \hspace*{20mm} (c) \\ \centering
\includegraphics[trim=0mm 0mm 5mm 0mm, clip, width=0.49\columnwidth]{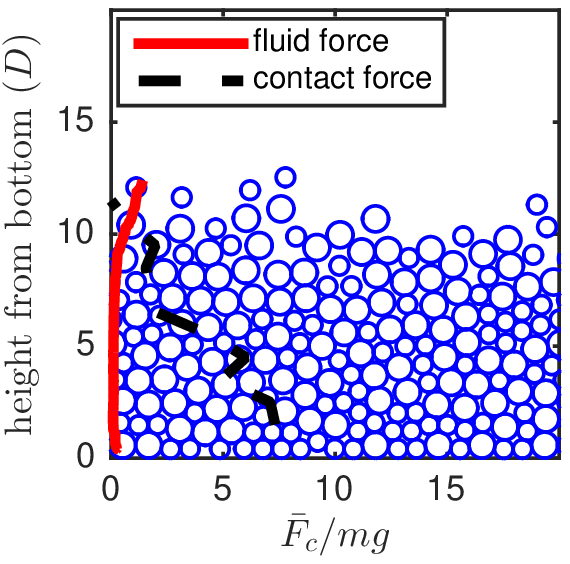}
\caption{(Color online) Profiles of the layer-averaged instantaneous fluid drag force (red lines) and the local pressure (black dashed lines) for $b=4$ and $\Gamma=0.25$, with (a) $\Theta=0.2$, (b) 0.35, and (c) 0.4.}
\label{fig:grain-stress}
\end{figure}

In Figure~\ref{fig:grain-stress}, we show typical force profiles during protocol A for grains under three different conditions. All panels have $b=4$ and $\Gamma=0.25$, but with varying $\Theta$. Solid red curves show the average force exerted by the fluid on the grains at a particular height. Black dashed curves show the average pressure due to grain-grain contacts as a function of height. To calculate the pressure, we first calculate the force moment tensor $M^i_{\alpha\beta}$ at particle $i$ by summing over the particles $j$ that contact particle $i$ to obtain $M^i_{\alpha\beta}=\frac{1}{R_i} \sum_j  F_\alpha^{ij}r_\beta^{ij}$, where $R_i$ is the particle radius, $\alpha$ and $\beta$ represent components, and $r_\beta^{ij}$ represents the $\beta$ component of the branch vector connecting the center of particle $i$ with the point of contact with particle $j$. The mean of the eigenvalues of this tensor provides a grain-scale estimate of the local pressure, which we then average over the horizontal direction to obtain the average contact force $\bar{F}_c$. We plot $\bar{F}_c$ in Fig.\ref{fig:grain-stress} as a function of height from the lower boundary as a black dashed line, which has roughly slope 1 in all plots. Thus, the contact forces below the surface are dominated by the weight of grains above a particular layer. Figure~\ref{fig:grain-stress}(a) shows $\Theta=0.2$, which is below both $\Theta_0$ and $\Theta_c$, so grains are not moving. Figure~\ref{fig:grain-stress}(b) shows $\Theta=0.35$, so the flow is metastable and grains will eventually come to a stop. Figure~\ref{fig:grain-stress}(c) shows $\Theta=0.4$, so grains continue to move indefinitely.

\subsection{\label{sec:protocol-B}Protocol B: Static-to-mobile transition}

For $\Theta>\Theta_c$, a sufficiently large perturbation will lead to
sustained bed motion. At high $Re_p$, this motion never begins for
$\Theta<\Theta_0$. But as the red crosses in Fig.~\ref{fig:phase-diagrams}
show, stable motion is not always initiated at these minimum values, but
depends on the particular arrangement of grains in the bed. We show that
sustained grain motion is consistent with Weibullian weakest link
statistics~\cite{weibull1939,weibull1951} and that failure events are
always initiated at $\Theta_c$ for low $Re_p$ and $\Theta_0$ for high $Re_p$ for sufficiently large
systems.

\begin{figure}
\raggedright (a) \\ 
\centering \includegraphics[trim=0mm 0mm 3mm 0mm, clip, width=\columnwidth]{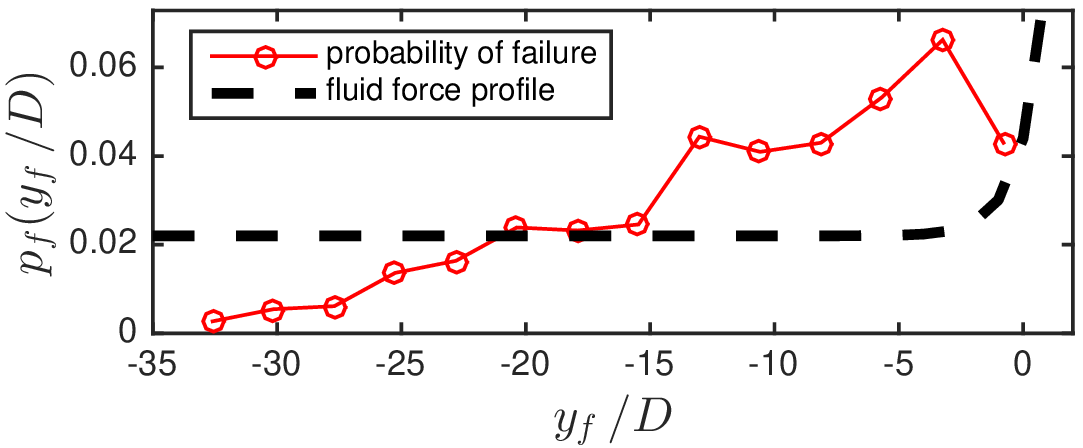}
\\ \raggedright (b) \\ 
\centering \includegraphics[trim=0mm 0mm 3mm 0mm, clip, width=\columnwidth]{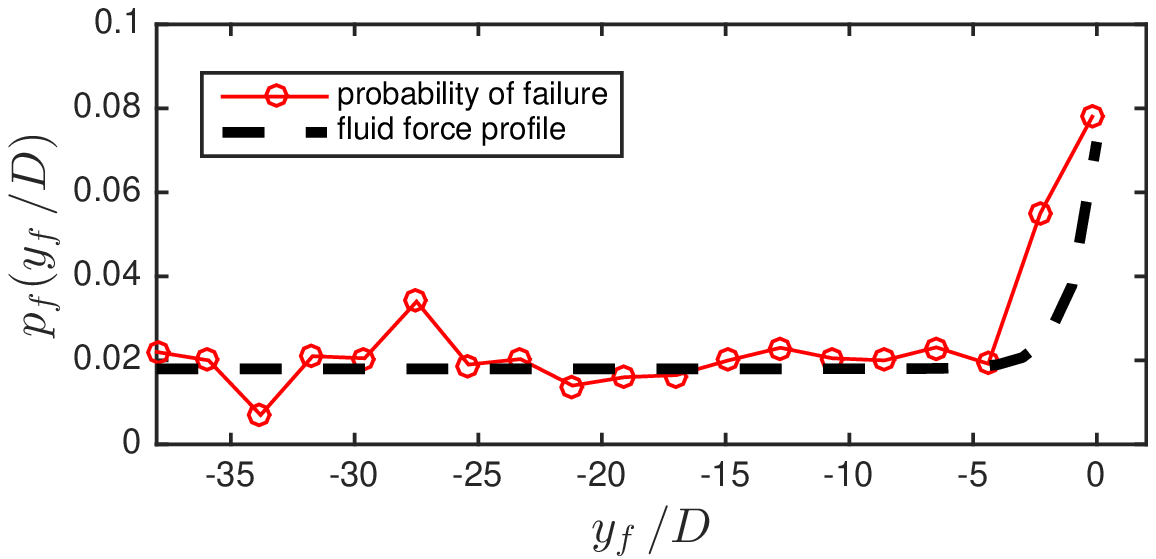}
\\ \raggedright (c) \\ 
\centering \includegraphics[trim=0mm 0mm 3mm 0mm, clip, width=\columnwidth]{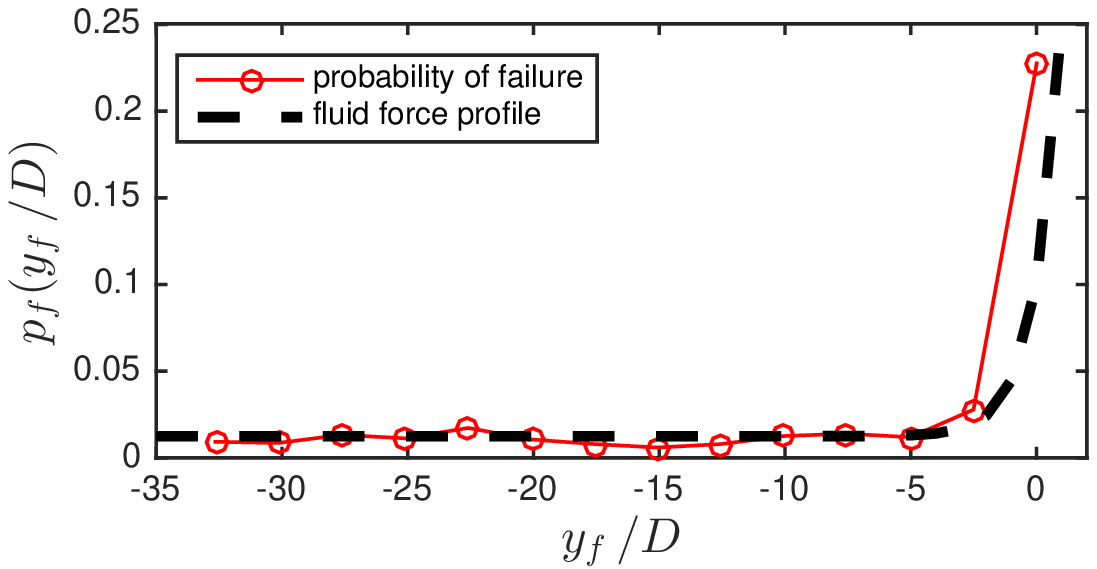}
\caption{(Color online) The probability $p_f$ that bed failure occurs at depth $y_f$ (where $y_f=0$ is the top of the bed) is proportional to the local fluid force for (a) $b=2$ and $\Gamma=0.1$, (b) $b=4$ and $\Gamma=0.25$, and (c) $b=6$ and $\Gamma=0.5$. The bed failure depth is determined as the depth of the particle whose acceleration first exceeds $a_{\rm thresh}$ (or a weighted average of depths, if multiple grains meet this condition simultaneously). The agreement between $p_f(y_f)$ and the local fluid force is good for $b=4$ and 6 cases, but there is a deviation for the $b=2$ case, such that $p_f$ is smaller than expected near the lower boundary and bed surface. One explanation for this is that failure events are less localized, causing the weighted average of $a_i>a_{\rm thresh}$ to be more likely in the middle of the system.}
\label{fig:failure}
\end{figure}

First, we note that although the first large-scale motion of the
grains always occurs at the top layer of grains, failure events do not
always originate there. If we measure the depth of the particle whose
acceleration first exceeds $a_{\rm thresh}$ at $\Theta_f$, we find
that it can occur at any depth $y_f$ below the bed surface, with a probability
distribution $p_f$ that is proportional to the local applied fluid
force as shown in Fig.~\ref{fig:failure}. Essentially, the ratio of the probability of failure at the surface to the probability of failure in the bed is roughly given by the ratio between the fluid force at the top of the bed to that inside the bed $v_0/v_b$, as given in Table~\ref{tbl:b-dependence}. Thus, while large-scale particle motion always begins at the surface of the bed, this motion is often correlated to small particle rearrangements that occur deep in the bed.

\begin{figure*}[ht!]
\raggedright (a) \hspace*{70mm} (b) \hspace*{45mm} (c)\\ 
\includegraphics[trim=0mm 0mm 0mm 0mm, clip, width=0.87\columnwidth]{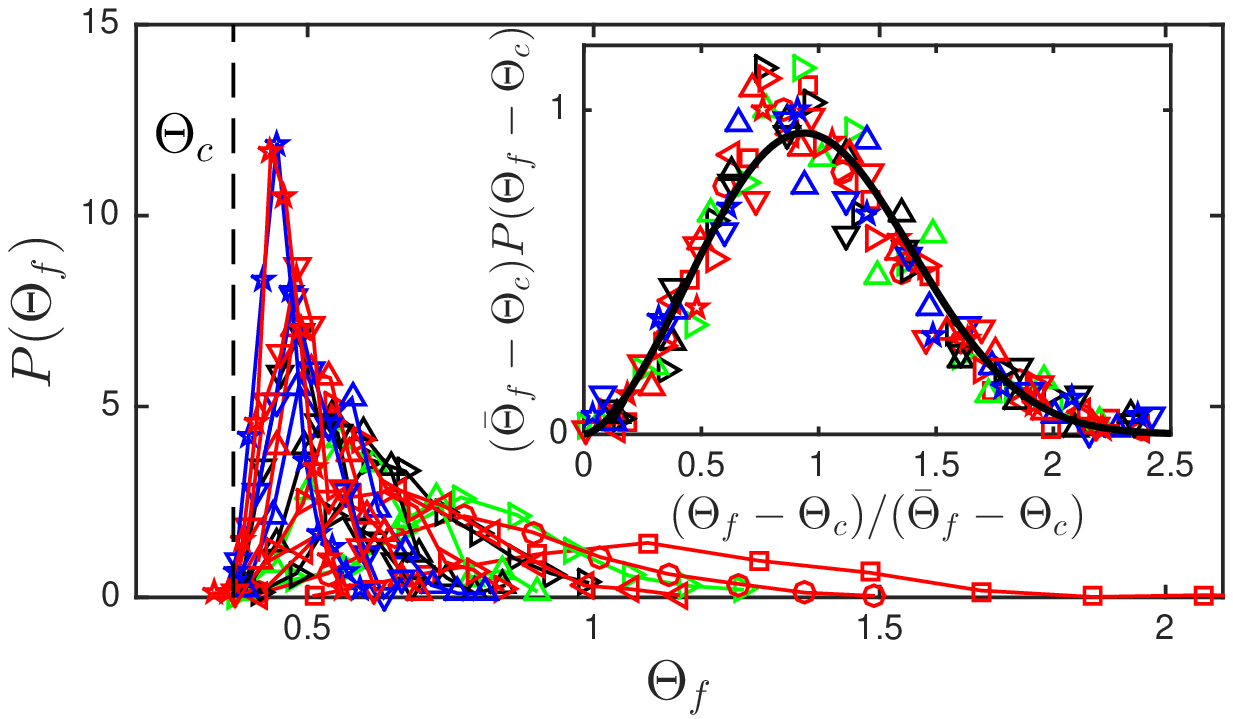}
\includegraphics[trim=0mm 0mm 0mm 3mm, clip, width=0.57\columnwidth]{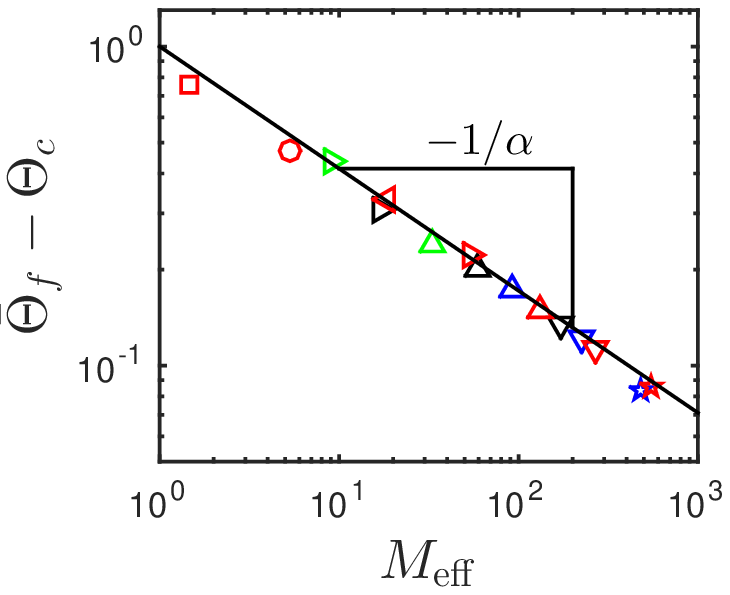}
\includegraphics[trim=0mm 0mm 5mm 0mm, clip, width=0.51\columnwidth]{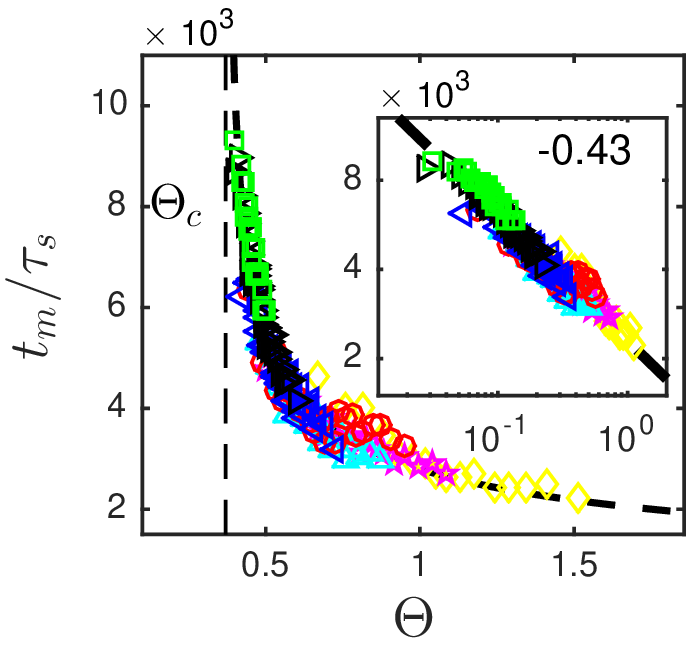}
\\ \raggedright (d) \hspace*{70mm} (e) \hspace*{45mm} (f)\\ 
\includegraphics[trim=0mm 0mm 0mm 0mm, clip, width=0.87\columnwidth]{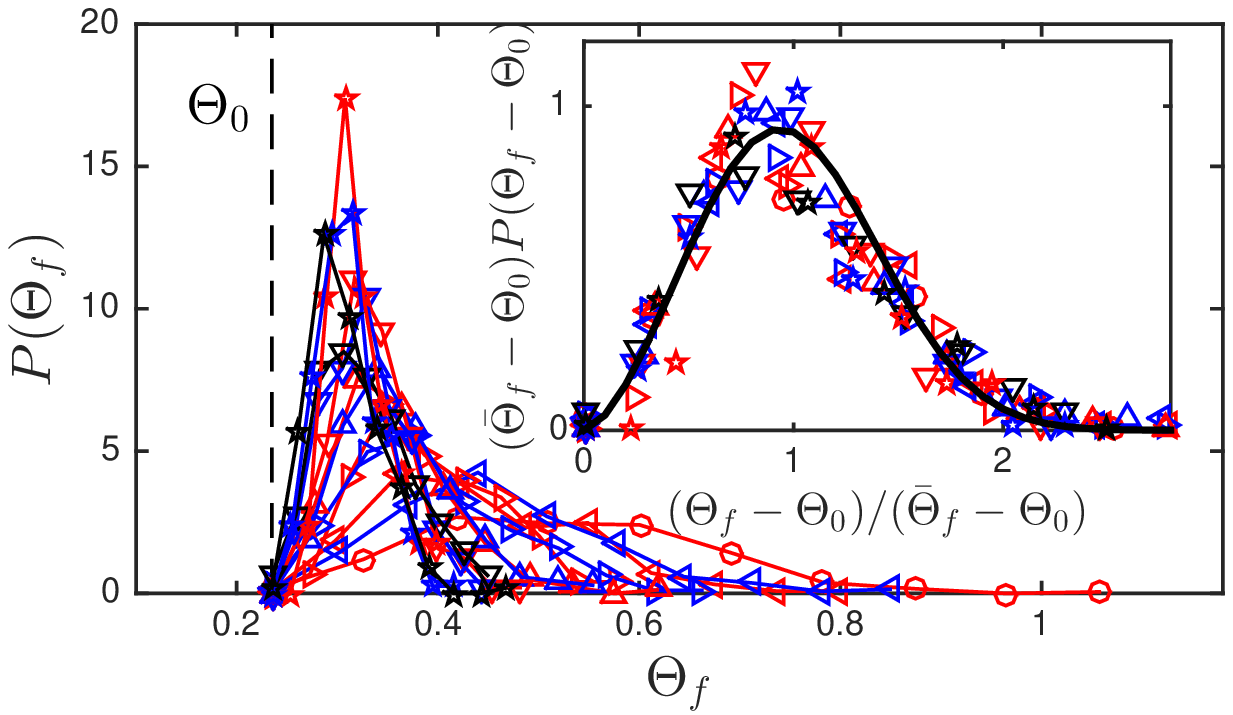}
\includegraphics[trim=0mm 0mm 0mm 3mm, clip, width=0.57\columnwidth]{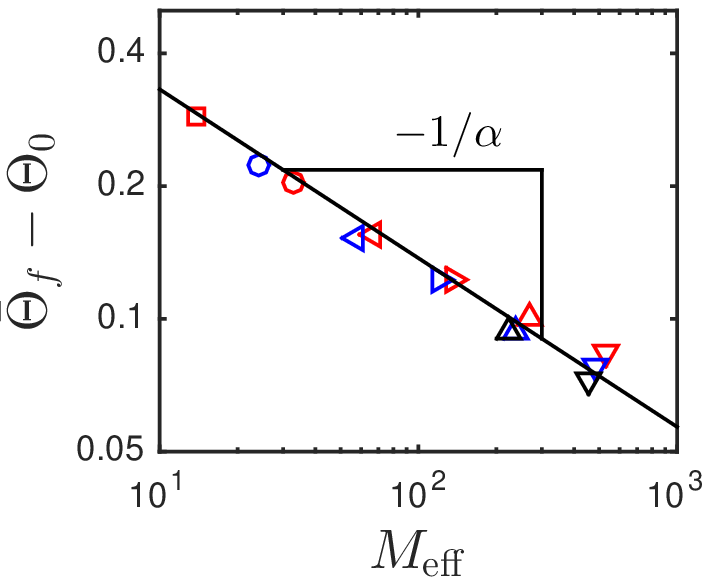}
\includegraphics[trim=0mm 0mm 5mm 0mm, clip, width=0.51\columnwidth]{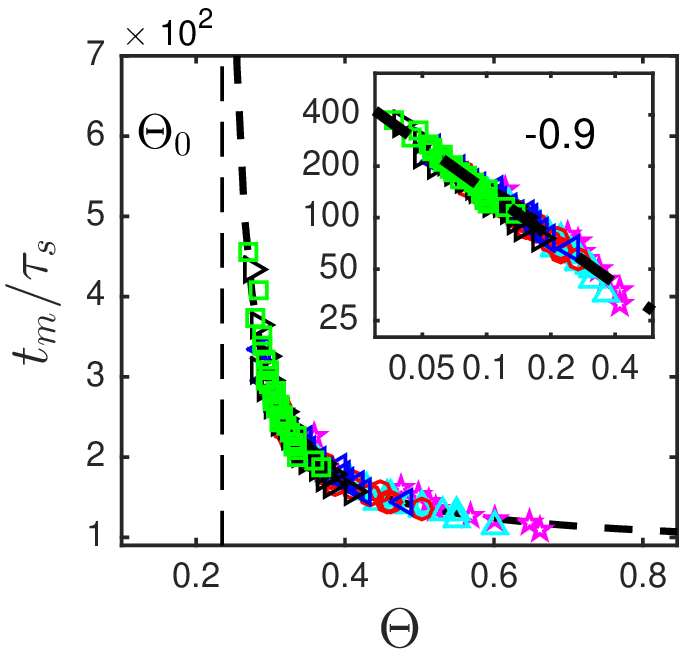}
\caption{(Color online) Onset of bed motion is governed by Weibullian
weakest-link statistics. (a)-(c) correspond to low $Re_p$ with
$\Gamma=0.25$ and (d)-(f) correspond to high $Re_p$ with
$\Gamma=0.1$. (a) and (d) show the probability distribution of the Shields number at bed failure $\Theta_f$ for many different
system sizes ($W/D$ from 3.125 to 200, $N$ from 25 to 1600, and
$WN/D$ from 8 to 80). The dashed vertical lines define (a) $\Theta_c$
and (d) $\Theta_0$. The insets show that $P(\Theta_f)$ collapses
when rescaled by $\bar{\Theta}_f-\Theta_c$, where $\bar{\Theta}_f$
is the mean of each distribution. The solid line is a Weibull
distribution with shape parameter $\alpha\approx 2.6$. (b) and (e)
indicate that $\bar{\Theta}_f-\Theta_c$ for low $Re_p$ and
$\bar{\Theta}_f-\Theta_0$ for high $Re_p$ scale with the effective
system size $M_{\rm eff}^{-1/\alpha}$. (c)
and (f) show that $t_m$, the mobilization time after bed failure,
diverges near $\Theta_c$ and $\Theta_0$, respectively, independent
of system size. The insets show a logarithmic plot of $t_m-t_{m,0}$
versus (c) $\Theta-\Theta_c$ and (f) $\Theta-\Theta_0$. The dashed
lines show $t_m-t_{m,0} \propto (\Theta-\Theta_c)^{-0.43}$ (c) and
$t_m-t_{m,0} \propto (\Theta-\Theta_c)^{-0.9}$ (f), and the thin
vertical dashed line indicates $\Theta_c$. Symbols
(\textcolor{yellow}{$\diamond$}, \textcolor{magenta}{$\star$},
\textcolor{cyan}{$\triangle$}, \textcolor{red}{$\circ$},
\textcolor{blue}{$\triangleleft$}, $\triangleright$,
\textcolor{green}{$\square$}) correspond to different values of $N$
($25$, $50$, $100$, $200$, $400$, $800$, and $1600$, respectively) with varying
$W/D$. Each data point represents an average of 20 simulations.}
\label{fig:Weibull-scaling}
\end{figure*}

Figure~\ref{fig:Weibull-scaling}(a) and (d) show that the probability
distribution $P(\Theta_f)$ approaches $\delta(\Theta_f-\Theta_c)$ for
low $Re_p$ and $\delta(\Theta_f-\Theta_0)$ for high $Re_p$ in the
large system limit, where we varied both $N$ and $W$ to change the
system size. The insets show that the distributions $P(\Theta_f)$
collapse when rescaled by their mean values. The system-size
dependence suggests a ``weakest link'' picture: at a given value of excess stress above $\Theta_c$ or $\Theta_0$ (for low and high $Re_p$, respectively), there is a better chance of finding a sufficiently weak local
grain arrangement somewhere in a larger system. If we consider the bed
to be a composite system of $M$ \textit{uncorrelated} subsystems that
fails if any of the subsystems fail, the cumulative distribution
$C_M(\Theta)$ for failure is related to that of a single subsystem
$C(\Theta)$ by
\begin{equation}
1-C_M(\Theta)=\left[1-C(\Theta)\right]^M.
\label{eqn:Weibull}
\end{equation} 
If we assume a Weibull distribution~\cite{weibull1939,weibull1951,franklin2014}
\begin{equation}
C(\Theta)=1-\exp\left[\left(\frac{\Theta-\Theta_c}{\beta}\right)^{\alpha}\right],
\label{eqn:Weibull-CDF}
\end{equation}
then $C_M(\Theta)$ will have the same form with $\alpha_M=\alpha$ and $\beta_M=\beta M^{-1/\alpha}$. Figure~\ref{fig:Weibull-scaling}(a) and (d) show that $P(\Theta_f)=dC/d\Theta_f$ does indeed obey a Weibull distribution with shape parameter $\alpha\approx 2.6$.

Thus, if Eq.~\eqref{eqn:Weibull-CDF} fits the data, and if $\alpha$ is constant but $\beta\propto M^{-1/\alpha}$ as $M$ is varied, then Eq.~\eqref{eqn:Weibull} applies, and global failure is caused by the failure of a single member of a collection of uncorrelated subsystems. Figure~\ref{fig:Weibull-scaling}(b) and (e) confirm this, showing that the mean Shields number at flow onset $\bar{\Theta}_f$ scales as $(\bar{\Theta}_f - \Theta_c) \propto M_{\rm eff} ^{-1/\alpha}$ for small $Re_p$ and $(\bar{\Theta}_f - \Theta_0) \propto M_{\rm eff} ^{-1/\alpha}$ for large $Re_p$, where $M_{\rm eff}=W_{\rm eff} H_{\rm eff} /D^2$ is the effective system size. This scaling means that larger systems are more likely to fail near $\Theta_c$ or $\Theta_0$ for small and large $Re_p$, respectively. However, systems that fail near these minimum values are very slow to become fully mobilized. To demonstrate this, we consider the mobilization time $t_m$, defined as the time between the initial force imbalance that leads to failure and the time when $\bar{v}_g$ reaches its asymptotic value. As shown in Fig.~\ref{fig:Weibull-scaling}(c) and (f), this time scale also diverges as $\Theta\rightarrow\Theta_c$ for low $Re_p$ and $\Theta\rightarrow\Theta_0$ for high $Re_p$, and is independent of system size.

To calculate $M_{\rm eff}$, we determine $W_{\rm eff}$ and $H_{\rm eff}$ as follows. Since the vertical symmetry is broken by the fluid forcing profile, we calculate $H_{\rm eff}$ by integrating the probability of failure over the depth of the system, which is equal to the fluid force profile. That is, $H_{\rm eff}$ is the integral of the profiles shown in Fig.~\ref{fig:failure} over the depth of the bed, since the relevant system size for the Weibullian weakest link scaling has to do with the probability of local failure, and, as previously mentioned, particles near the surface cause failure with a likelihood that is greater than particles beneath the surface by a factor $v_0/v_b$. We find that this form for $H_{\rm eff}$ collapses the data for systems that are sufficiently large in the horizontal (periodic) direction. We also check the form of $H_{\rm eff}$ for a case where $b=2$ and $\Gamma=0.1$, such that $v_0/v_b\approx 1.97$, and we find it to collapse the data well according to the Weibullian scaling. The form of $H_{\rm eff}$ confirms our observation that surface grain motion can be initiated from deep beneath the surface, as it is also proportional to the local applied shear force.

The horizontal dimension is symmetric, but we find finite system-size effects when $W/D<\xi$, where $\xi$ is a horizontal correlation length that varies with $Re_p$. As shown in Fig.~\ref{fig:Weibull-scaling}, we find $M_{\rm eff} = 0.91 (\bar{\Theta}_f - \Theta_c)^{-\alpha}$ for large systems, where $H_{\rm eff}$ is calculated using the method described previously. However, when the horizontal dimension becomes small, we observe that $(\bar{\Theta}_f - \Theta_c)$ is larger than expected, which corresponds to an effective system width $W_{\rm eff}$ that is smaller than the real system width $W$.

\begin{figure}
\includegraphics[trim=0mm 0mm 0mm 0mm, clip, width=0.95\columnwidth]{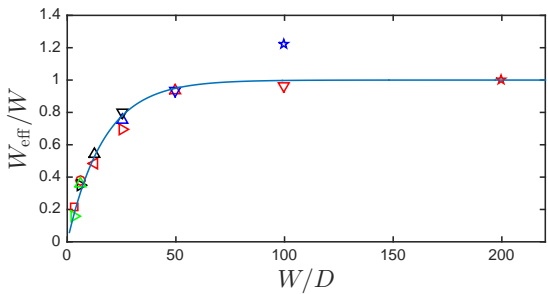}
\caption{(Color online) This plot shows the finite system size effects in the horizontal periodic direction (with $b=4$ and $\Gamma=0.25$, which are the same values for the low $Re_p$ data shown in Fig.~\ref{fig:Weibull-scaling}(a)-(c)). The horizontal axis is the system width in particle diameters, $W/D$. As we show in in Fig.~\ref{fig:Weibull-scaling}(a)-(c), we find $M_{\rm eff} = 0.91 (\bar{\Theta}_f - \Theta_c)^{-\alpha}$ for large systems. The vertical axis shows this quantity, $0.91(\bar{\Theta}_f - \Theta_c)^{-\alpha}$, divided by the product of the effective fill height $H_{\rm eff}/D$ and the width in particle diameters $W/D$. This shows a good collapse, and the fit line corresponds to $W_{\rm eff}/W = 1-\exp\left[-\frac{W}{\xi D}\right]$, with $\xi=16.7$, and 95\% confidence interval of roughly $15<\xi<20$.}
\label{fig:finite-size}
\end{figure}

If $M_{\rm eff} = 0.91 (\bar{\Theta}_f - \Theta_c)^{-\alpha}=H_{\rm eff} W_{\rm eff}/D^2$, and we assume that $W_{\rm eff}/W=\zeta (W/D)$, then we can write
\begin{equation}
\zeta(W/D) = 0.91 (\bar{\Theta}_f - \Theta_c)^{-\alpha} \frac{D^2}{H_{\rm eff} W}.
\label{eqn:finite-size}
\end{equation}
A plot of this quantity is shown in Fig.~\ref{fig:finite-size}, and the fit line corresponds to
\begin{equation}
\zeta(W/D) = 1-\exp\left[-\frac{W}{\xi D}\right].
\label{eqn:finite-size-2}
\end{equation}

As Fig.~\ref{fig:finite-size} shows, the form $W_{\rm eff}/W = 1-\exp\left[-\frac{W}{\xi D}\right]$ captures the finite size effects. We find a similar result for the the high $Re_p$ case in Fig.~\ref{fig:Weibull-scaling}(d)-(f), where $\Gamma=0.1$ and $b=4$, but with a smaller value of $\xi\approx 3$. The form of $W_{\rm eff}$ suggests a horizontal correlation length of roughly $\xi$, which is larger for low $Re_p$ ($\xi\approx 17$) than for high $Re_p$ ($\xi\approx 3$).

\section{\label{sec:summary}Summary}

In summary, we performed numerical simulations of a granular bed
subjected to a simple fluid shear flow to understand general features
of the initiation and cessation of grain motion. The critical Shields
number for the onset of grain motion $\Theta_c(Re_p)$ from the
simulations is consistent with the behavior from a large body of
experimental results~\cite{buffington1997}. At low $Re_p$,
$\Theta_c(Re_p)$ separates mobile and static beds, but at high $Re_p$,
we observe significant hysteresis as a consequence of particle
inertia. We find that the onset of grain motion is directly connected to the
packing structure, even deep in the bed where there is a weak
but nonzero fluid stress~\cite{rose1945,beavers1967,scheidegger1974}. Our
results from this simple model clarify the essential physics
governing the transition between mobile and static beds. In future
work, additional effects such as turbulent flow, inter-grain friction,
and nontrivial particle shape can be added one-by-one to determine
their distinct effects on the onset and cessation of grain motion.

\begin{acknowledgments}
This work was supported by the US Army Research Office under Grant No.~W911NF-14-1-0005.
\end{acknowledgments}

\bibliography{erosion_sim}

\end{document}